\documentclass[12pt]{article}
\usepackage{pdflscape}
\usepackage{amsmath}
\usepackage{amssymb}
\usepackage{dcolumn}
\usepackage{bm}
\usepackage{color}
\usepackage{epsfig}
\usepackage{amsfonts}
\usepackage{graphicx}
\usepackage{subfigure}
\usepackage{dcolumn}
\usepackage{jheppub_kim}
\usepackage{enumerate}

\usepackage{amsmath}  
	
\newcommand{\bse}{\begin{subequations}}
	\newcommand{\ese}{\end{subequations}}
\newcommand{\be}{\begin{equation}}
\newcommand{\ee}{\end{equation}}
\newcommand{\bea}{\begin{eqnarray}}
\newcommand{\eea}{\end{eqnarray}}
\newcommand{\ba}{\begin{array}}
	\newcommand{\ea}{\end{array}}

\usepackage{color}

\usepackage[normalem]{ulem}  

\newcommand{\non}{\nonumber\\}

\begin{document}
\title{Holographic nuclear matter in presence of chemical potential at finite temperature}

\author[a]{Farideh Kazemian,}
\author[a]{Behnam Pourhassan}

\affiliation[a]{School of Physics, Damghan University, P. O. Box 3671641167, Damghan, Iran.}

\emailAdd{fkazemian.144@gmail.com}
\emailAdd{b.pourhassan@du.ac.ir}
\abstract{
We study the dense nuclear matter within the holographic Sakai-Sugimoto model. The nuclear matter is described via instantons in bulk, whose size has the new temperature dependency. Then, properties of nuclear matter have been studied for different temperatures. For example, free energy and baryon density are examined for different temperature values. Also, nuclear matter properties, like the speed of sound and connection of nuclear matter and quark matter, are discussed. As we have increased the temperature, the phase transition value has been changed from baryon to quark, which is non-physical. Also, we can see continuity between the baryonic and quark phases. Speed of  sound at the low chemical potential in the presence of temperature is different, but when the potential increases, the effect of temperature on the speed of sound will remain unchanged. In the phase diagram, for significant chemical potential, chiral symmetry is restored and provides all things that we have the realistic model.
}


\maketitle

\section{Introduction}
Calculations in quantum chromodynamics (QCD) for the matter at low temperature and considerable chemical potential are complicated. Two important examples of matter in this regime are nuclear matter inside compact stars (including neutrons and protons in the simplest combinations with possible Cooper pairing), and deconfined matter (including three-flavor quarks). An important property of compact stars is their rich phase diagram where differs them from ordinary matter in the condensed matter systems. They include both nuclear and deconfined matters. However, studying both phases in one field theory model is very difficult. Also, these systems are strongly coupled, and study of them with conventional lattice QCD models is challenging, see \cite{Aarts:2015tyj} for recent progress and the other models like Nambu-Jona-Lasinio (NJL) \cite{Nambu:1961fr} and quark-meson \cite{Schaefer:2007pw} have problems, they can be beneficial to get some insight into quark matter. Still, they usually do not include nuclear matter.\\
The original version of gauge/string correspondence is $N=4$ supersymmetric Yang-Mills theory for dual field theory and type IIB string theory for gravity. This exciting method is helps study strongly coupled gauge theory, especially for zero or small baryon chemical potential. Important studies include  application to the heavy-ion collision.\\
In this paper, we use Sakai-Sugimoto (SS) model and study QCD properties. Already, the full spectrum of the top-down $D4$-$D8$ brane system of Sakai-Sugimoto model calculated by Ref. \cite{2010.00025}. We can account for a single model for nuclear and quark matter, where the parameters of the model are the 't Hooft coupling $\lambda$, the Kaluza-Klein mass $M_{KK}$, and the asymptotic separations of $D8$ and $\bar{D}8$ branes which denoted by $L$. In this model, fundamental chiral fermions have been introduced through $N_{f}$ $D8$ and $\Bar{D}8$ branes. These are in the background of $N_{c}$ $D4$ branes \cite{Witten:1998qj}, $N_{c}$ and $N_{f}$ are respectively numbers of color and flavor. Although, these are another works where Sakai-Sugimoto considered, so effect of chemical potential on the nuclear force by using the string theory already studied by Ref. \cite{pourhasan:2020}.\\
We now work in the  $N_{c}>>N_{f}$ limit, that this limit caused to neglect effects of flavor branes on the background geometry, and also we consider the chiral limit when $D8$ and $\Bar{D}8$ branes are not connected. however, you can see Ref. \cite{Schmitt:2020} for discussing non-zero quark mass.  Baryons are studied as $D4$-branes wrapped around the 4-sphere of the background geometry in the Sakai-Sugimoto model \cite{hashimoto:2008}, here, this is the same to instanton with the non-zero topological charge on the connected flavor branes of the model \cite{Sakai:2005yt}, baryonic matter at first, non-zero density, and temperature was considered in a point-like approximation of the instantons on the flavor branes \cite{Bergman:2007pm}. The instanton has been further developed including finite-size effects \cite{Li:2015uea}. It has been studied by Ref. \cite{Preis:2016fsp}, but here we extend \cite{bitaghsir:2018kfa} and consider a new profile for instanton which depends on the temperature. Indeed, the temperature dependence was already studied in
Ref. \cite{Preis:2016fsp} where the numerical analysis was only carried
out in the zero temperature regime. Now, we would like to do numerical analysis of the finite temperature.\\
We are working in the decomapctified limit of the SS model\cite{Antonyan, Davis2007, FAA 2013} and this is the
limit where the asymptotic separation of D8 branes ($L$) is small compared
to the compactified radius of the extra dimension. In another word, this limit have been
provided by fixing $L$ and choosing a very large radius. Since a large radius
corresponds to a very small Kaluza-Klein mass, the critical temperature
for deconfinement goes to zero in this limit then this limit is suitable for
studying the phase transitions, which now depends on the baryon chemical
potential and temperature.\\
The holographic description of quark matter in this setup is giving by disconnected flavor branes. In Ref. \cite{bitaghsir:2018kfa}, it was shown that nuclear and quark matter could be continuously transformed into each other when we have considered  the interaction between instantons at zero temperature. Our work is an extension of Ref. \cite{Preis:2016gvf}, by accounting for the interactions of the instantons in bulk. It has been done by using the exact two-instanton solution in flat space, which is a special case of the general Atiyah-Drinfeld-Hitchin-Manin (ADHM) construction \cite{Bergman:2007pm}, and which has been discussed previously in the context of the SS model \cite{Hashimoto:2009ys} to study the nucleon-nucleon interaction \cite{Atiyah:1978ri,Kim:2008iy}. As we shall see, the instanton interactions are crucial for this observation. There are various other, in some aspects complementary, approximations to many-baryon phases in the Sakai-Sugimoto model, see for instance Refs. \cite{Kim:2008iy,Bolognesi:2014dja}. One of them is based on a homogeneous ansatz for the gauge fields in the bulk \cite{Rozali:2007rx}, which is expected to yield a better approximation for large densities, which is less transparent from a physical point of view because it is not built from single instantons.\\
This paper is organized as follows: In section 2, we introduce DBI and CS actions, section 3 including deformation parameters and depending on the temperature.
In section 4, we explained results like a phase diagram, connecting two phases and the speed of sound. Finally, in section 5, we give a conclusion and
summary of results.
\section{Geometry}
Our calculation has been started with the following Dirac-Born-Infeld and Chern-Simons action,
\begin{equation}
S=S_{DBI}+S_{CS},
\end{equation}
where $S_{DBI}$ is given as
\begin{equation}
S_{\rm DBI} = 2T_8 V_4\int_0^{1/T} d\tau \int d^3X \int_{u_c}^\infty du \, e^{-\Phi}\sqrt{{\rm det}(g+2\pi\alpha' {\cal F})},
\end{equation}
where $\alpha'$ is related
to string length $\ell_{s} $ via $\alpha'=\ell_{s}^{2}$,
$T_{8}=\frac{1}{(2\pi)^{8}\ell_{s}^{9}} $ is $D8$-brane tension,
$V_{4}=\frac{8\pi^{2}}{3}$ is the volume of the 4-sphere. One should notice that $u_{c}$ is the location of the tip of the connected flavor branes $D8$ and $\Bar{D}8$. The dilaton field is given via $ e^{\Phi}=g_{s}\left(\frac{U}{R}\right)^{\frac{3}{4}} $, where $R$ is the curvature radius and $g_{s}$ is the string coupling. Also,
\begin{equation}
u=\frac{U}{M_{KK}R^{3}},
\end{equation}
is holographic coordinate which is dimensionless quantity.
The integral is taken over the position space, imaginary time $\tau$ with the temperature $T$ and over the holographic coordinate $u$, also $g$ is the induced metric on the flavor branes. The detailed form of its induced metric $g$ on the $D8$ brane, abelian and non-abelian field strengths can be found in Ref. \cite{bitaghsir:2018kfa}. Also,
\begin{equation}
{\cal F}_{\mu\nu} = \hat{F}_{\mu\nu} + F_{\mu\nu} \ \, , \qquad F_{\mu\nu}= F_{\mu\nu}^a \sigma_a,
\end{equation}
with the Pauli matrices $\sigma_{a}$ and $\mu,\nu=0,1,2,3,u$. In our ansatz, the only non-zero abelian field strength is $\hat{F}_{0u}$ and non-abelian field strengths are $F_{iu}$, and $F_{ij}$. All other field strengths are set to zero.\\
By using the mentioned ansatz, the CS contribution is given by,
\begin{equation}
S_{\rm CS} = \frac{N_c}{4\pi^2}\int_{0}^{1/T} d\tau \int d^3X \int_{U_{c}}^\infty du \hat{A}_{0} Tr[F_{ij}F_{kU}]\epsilon_{ijk}.
\end{equation}
Therefore, we can obtain
\begin{equation}
{\cal L} = u^{5/2}\sqrt{(1+u^3f_T x_4^{\prime2}-\hat{a}_0^{\prime2}+g_1)(1+g_2)}-\hat{a}_0 n_I q(u),
\end{equation}
where the prime denotes derivative with respect to $u$ and $ n_{I} $ represent the instanton density per flavor . Also, $f_{T}=1-u_{T}^{3}/u^{3}$ with $u_{T}=(\frac{4\pi T}{3M_{KK}})^2$. The instanton density is given by $n_{I}$, and $\hat{a}_{0}$ is denoted the abelian gauge field with boundary condition so $\hat{a}_{0}(\infty)=\mu$. Also the embedding function of the flavor branes has been shown by $x_{4}(u)$ so the boundary conditions are $x_{4}(u_{c})=0$ and $x_{4}(\infty)=LM_{KK}/2$.
The functions $g_{1}$, $g_{2}$ and $q$ in Ref. \cite{bitaghsir:2018kfa} are defined by the following relations
\begin{equation}
g_{1}\equiv\frac{f_{T}n_{I}}{3\gamma}\frac{\partial z}{\partial u}q(u),\ \ \ \ \ \ \ \ \ \ g_{2}\equiv\frac{\gamma n_{I}}{3 u^{3}}\frac{\partial u}{\partial z}q(u),
\end{equation}
where $u=(u_{c}^3+u_{c}z^2)^{1/3}$ shows the relation between $z$ and $u$, also
\begin{equation}
q(u)=2\frac{\partial z}{\partial u}D(z),\ \ \ \ \ \ \ \ \ \ \ \ \ \ \ D(z)=(1-p)D_{0}(z)+\frac{p}{2}D_{int}(d,z),
\end{equation}
where the function $D(z)$ is instanton profile on the flavor branes which is depends on the lattice parameter $p$. So, the solution for the single instanton is given by
\be
D_{0}(z)=\frac{3\rho^{4}}{4(\rho^{2}+z^{2})^{5/2}},
\ee
where $\rho$ is the width in the holographic direction. We can control the interaction between the instantons of this model by introducing the instanton interaction as
\be
D_{int}(z)=\frac{3\sqrt{2}\rho^{8}}{4}\frac{h_{1}S_{1}+h_{2}S_{2}}{(a^{2}+b)^{5/2}b^{2}}.
\ee
This result has been derived by Ref. \cite{bitaghsir:2018kfa} from ADHM approach, all parameters, $ h_{1},S_{1},h_{2},S_{2}, a$ and $ b $ are defined in the appendix A. We have also introduced $ d $ as the overlap parameter
\begin{equation*}
d=\frac{\gamma}{\rho}(\frac{6\pi^{4}r}{\lambda^{2}n_{I}}),
\end{equation*}
where $\gamma$ is a deformation parameter. It is interpreted as a distance between two instantons that normalized by twice their spatial width; when $d\rightarrow0$, then instantons will overlap.\\
In the above equations, we have parameters $ p $ and $ r $ that all information about lattice structures are carried by them. For example, cubic, a body-centered cubic and a face-centered cubic crystal we have $ (p,r)=(6,1), (8,3\sqrt{3}/4)$ and $(12,\sqrt{2}) $
 respectively. For non-interacting approximation, we consider $ p=0 $.\\
 We have checked for other lattice configurations and saw that the results do not depend significantly on the lattice configurations, then we present our results only for cubic lattice $p=12,r=\sqrt{2}$.
\\
\section{Deformed instantons and thermodynamic quantities}
The main novelty of this paper is to study the new temperature dependency of the instantons shape. We introduced instantons in our configuration setup. In the shape of instanton, $\rho$ is the width in the holographic direction while $\rho/\gamma$ is the spatial width; as a result $\gamma$ is a deformation parameter that characterizes the deviation of the instanton from SO(4) symmetry. For the single instanton at large t'Hooft coupling, the leading-order expression has been considered for deconfined geometry \cite{Preis:2016fsp}:
\begin{equation}
(F_{iz})^{2}_{(1)}=\frac{12(\rho/\gamma)^{4}}{\gamma^2[x^2+(z/\gamma)^2+(\rho/\gamma)^{2}]^{4}},
\end{equation}
where the subscript "(1)" indicated the single-instanton solution, also $x^{2}=x_{1}^{2}+x_{2}^{2}+x_{3}^{2} $ and $z$ indicates the holographic coordinate where $z\epsilon [-\infty,\infty]$. Moreover, parameters $\rho$ and $\gamma$ that described the shape of the instanton are given by the following expressions,
\begin{eqnarray}\label{shapeT}
\rho&=&\frac{\rho_{0}u_{c}^{3/4}}{\sqrt{\lambda}}\left[\frac{f_{T}(u_{c})}{\beta_{T}(u_{c})}\right]^{1/4}\sqrt{\alpha_{T}(u_{c})},\nonumber\\ \gamma&=&\frac{3\gamma_{0}u_{c}^{3/2}}{2}\sqrt{\alpha_{T}(u_{c})}.
\end{eqnarray}
Also
\begin{eqnarray}
\alpha_{T}(u_{c})&\equiv& 1-\frac{5u_{T}^{3}}{8u_{c}^{3}},\nonumber\\
\beta_{T}(u_{c})&\equiv& 1-\frac{u_{T}^{3}}{8u_{c}^{3}}-\frac{5u_{T}^{6}}{16u_{c}^{6}}.
\end{eqnarray}
At zero temperature we have $f_{T}(u_{c})=\alpha_{T}(u_{c})=\beta_{T}(u_{c})=1 $. Also $\rho_{0} $ and $ \gamma_{0} $ are free parameters which have been introduced in Refs. \cite{Preis:2016fsp} and \cite{Preis:2016gvf}. Our purpose is to investigate how important thermodynamic properties of the given system, such as the phase diagram, change. In Table 1, as well as Ref. \cite{Schmitt:2015}, we have obtained the physical dimensionful quantities from their dimensionless counterparts $t,\mu, n_{I}, M$. We have abbreviated $\lambda_{0}=\frac{\lambda}{4\pi}$ and $M_{KK}$ is Kaluza-Klein mass.\\

\begin{table*}
\begin{center}
\begin{tabular}{|c|c|c|c|}
\hline
\rule[-0.5ex]{0em}{2ex}
 temperature &  chemical potential &  number density &
 constituent mass
\\[1ex]
\hline
\rule[-0.5ex]{0em}{5ex}
$M_{\rm KK} t$ & $\lambda_0 M_{\rm KK}\mu $ & $\displaystyle{\frac{N_fN_cM_{\rm KK}^3\lambda_0^2}{6\pi^2}n}$ &
$\lambda_0 M_{\rm KK}M_q$
\\[1ex] \hline
\end{tabular}
\end{center}
\caption{The physical, dimensional quantities from their dimensionless quantities $t,\mu,n,M_{q}$, in the paper, we use capital letter for dimensionful and small letter for dimensionless; for example: $T=M_{kk}t$. }
\label{table1}
\end{table*}

Here, we present the main equations and formulas for computing thermodynamic quantities of the system. In next section we will use these equations to find these quantities by numeric techniques.
\begin{itemize}
\item{Free energy}\\
 The main quantity is the free energy of the system, one finds it as
 \begin{equation}
\Omega = \int_{u_c}^\infty du {\cal L},
\end{equation}
which is yields the following expression
\begin{equation}\label{free}
\Omega=\int_{u_c}^\infty du \left(u^{5/2} \eta (u) +\frac{\ell}{2}k-\mu n_{I}\right),
\end{equation}
where
\begin{equation}
\eta (u) =\sqrt{1+g_{1}}\sqrt{1+g_{2}-\frac{k^{2}}{u^{8}f_{T}}+\frac{(n_{I}Q)^2}{u^5}},
\end{equation}
where $k$ is an integration constant. \\

To minimize the free energy, $\Omega$ with respect to free parameters, at given $ T $ and $ \mu $, we calculate $ \frac{\partial \Omega}{\partial k} = \frac{\partial \Omega}{\partial n_I} = \frac{\partial \Omega}{\partial u_c} = 0$.
Therefore, one finds the three main stationarity equations as \\
\begin{subequations} \allowdisplaybreaks \label{minimOm}
\bea
\frac{\ell}{2} &=& \int_{u_c}^\infty du\, x_4' \, , \label{dk} \\[2ex]
\mu n_I &=&\int_{u_c}^\infty du \,u^{5/2}\left[\frac{g_1\zeta^{-1}+g_2\zeta}{2q}\left(q-\frac{d}{3}\frac{\partial q}{\partial d}\right) + \frac{\zeta n_I^2 Q}{u^5}\left(Q -\frac{d}{3}\frac{\partial Q}{\partial d}\right)\right]
\, , \label{dnI} \\[2ex]
 st &=& 2u_c^{7/2}+\int_{u_c}^\infty du \, u^{5/2} \left\{7 - \zeta\left[7(1+g_2)+2\frac{(n_I Q)^2}{u^5}+\frac{k^2}{u^8f_T}\right] +
 \right .
 \non[2ex]
&& \left .
\frac{g_1\zeta^{-1}+g_2\zeta}{2q}\left(5q\!-\!\frac{3d}{2}\frac{\partial q}{\partial d}
 +\frac{\rho}{2}\frac{\partial q}{\partial \rho}\right)-\frac{\zeta n_I^2Q}{u^5}\left(\frac{3d}{2}\frac{\partial Q}{\partial d}-\frac{\rho}{2} \frac{\partial Q}{\partial \rho}\right) \right\}
 \eea
\end{subequations}
 and
\begin{equation}
Q(u)=\int_{uc}^{u}dv q(v)
\end{equation}

\item{Entropy density}\\
By switching on the temperature in the system, the dimensionless entropy density is denoted by, $s$, which is entropy of the system. First, we give the derivation of it from the free energy, one knows that $ \Omega=\Omega(u_c,k,n_{I},\rho,\gamma,d)$ then,
\begin{equation}
s=-\frac{\partial \Omega}{\partial t}=-\frac{\partial \Omega}{\partial u_{c}}\frac{\partial u_{c}}{\partial t}-\frac{\partial \Omega}{\partial k}\frac{\partial k}{\partial t}-\frac{\partial \Omega}{\partial n_{I}}\frac{\partial n_{I}}{\partial t}-\frac{\partial \Omega}{\partial \rho}\frac{\partial \rho}{\partial t}-\frac{\partial\Omega}{\partial \gamma}\frac{\partial\gamma}{\partial t}-\frac{\partial \Omega}{\partial d}\frac{\partial d}{\partial t}-\frac{\partial \Omega}{\partial t}.
\label{entorpyN}
\end{equation}
The first three terms are zero because of stationary point, also we have,
\begin{equation}
\frac{\partial\Omega}{\partial \rho}=\int_{u_c}^\infty du \, u^{5/2} \left(\frac{1}{2q}(g_{1}\zeta^{-1}+g_{2}\zeta)\frac{\partial q}{\partial \rho}+\frac{n_{I}^{2}}{u^{5}}Q\frac{\partial Q}{\partial \rho}\zeta \right),
\end{equation}
\begin{equation}
\frac{\partial \Omega}{\partial \gamma}=\int_{u_c}^\infty du \, u^{5/2} \left(\frac{1}{2q}(g_{1}\xi^{-1}+g_{2}\zeta)\frac{\partial q}{\partial \gamma}+\frac{n_{I}^{2}}{u^{5}}Q\frac{\partial Q}{\partial \gamma}\zeta+\frac{1}{2}(\frac{g_{1}\zeta^{-1}+g_{2}\zeta}{\gamma})\right),
\end{equation}
and
\begin{equation}
\frac{\partial \Omega}{\partial d}=\int_{u_c}^\infty du \, u^{5/2} \left(\frac{1}{2q}(g_{1}\zeta^{-1}+g_{2}\zeta)\frac{\partial q}{\partial d}+\frac{n_{I}^{2}}{u^{5}}Q\frac{\partial Q}{\partial d}\zeta\right).
\end{equation}
Moreover,
\begin{equation}
\frac{\partial\Omega}{\partial t}=\frac{3u_{T}^{3}}{t}\int_{u_c}^\infty \frac{du}{\sqrt{u}f_{T}}\left(g_{1}\zeta^{-1}+\frac{k^{2}}{u^{8}f_{T}}\zeta\right).
\end{equation}
Finally, one can obtain,
\begin{align}
s&=\int_{u_c}^\infty du \left[\frac{3u_{T}^{3}}{t}\frac{1}{\sqrt{u}f_{T}}(g_{1}\zeta^{-1}+\frac{k^{2}}{u^{8}f_{T}}\zeta)+u^{5/2}\mathcal{X}\right],
\end{align}

where we defined
\begin{eqnarray}
\mathcal{X}&\equiv&\left(\frac{1}{2q}(g_{1}\zeta^{-1}
+g_{2}\zeta)\frac{dq}{d\rho}+\frac{n_{I}^{2}}{u^{5}}Q\frac{dQ}{d\rho}\zeta\right)\frac{d\rho}{dt}\nonumber\\
&+&\frac{1}{2}(\frac{g_{1}\xi^{-1}+g_{2}\zeta}{\gamma})\frac{d\gamma}{dt}+(\frac{g_{1}\zeta^{-1}
+g_{2}\zeta}{2q}\frac{dq}{dd}+\frac{n_{I}^{2}}{u^{5}}Q\frac{dQ}{dd}\xi)\frac{dd}{dt},
\end{eqnarray}
and
\begin{equation}\label{26}
\zeta=\frac{\sqrt{1+g_{1}}}{\sqrt{1+g_{2}-\frac{k^{2}}{u^{8}f_{T}}+\frac{(n_{I}Q)^{2}}{u^{5}}}}.
\end{equation}

\begin{figure}
\centering
\hbox{\includegraphics[width=.49\textwidth]{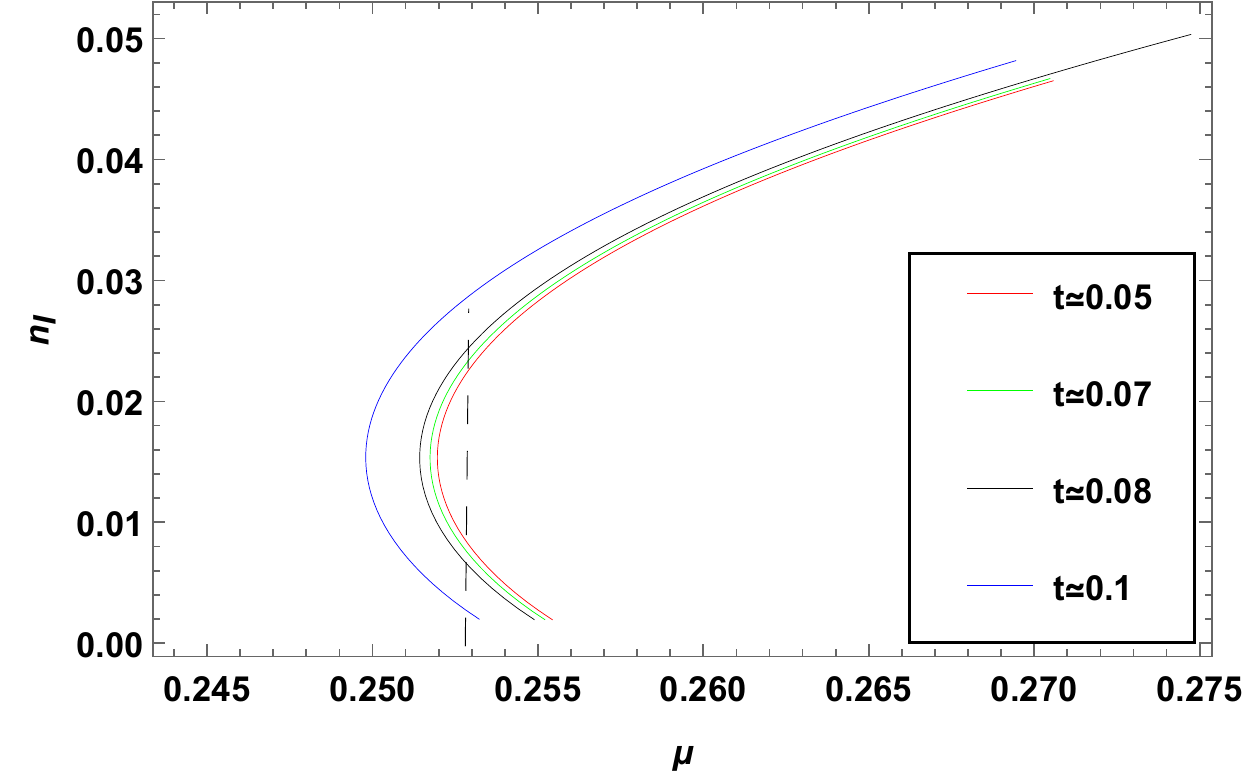}\includegraphics[width=.49\textwidth]{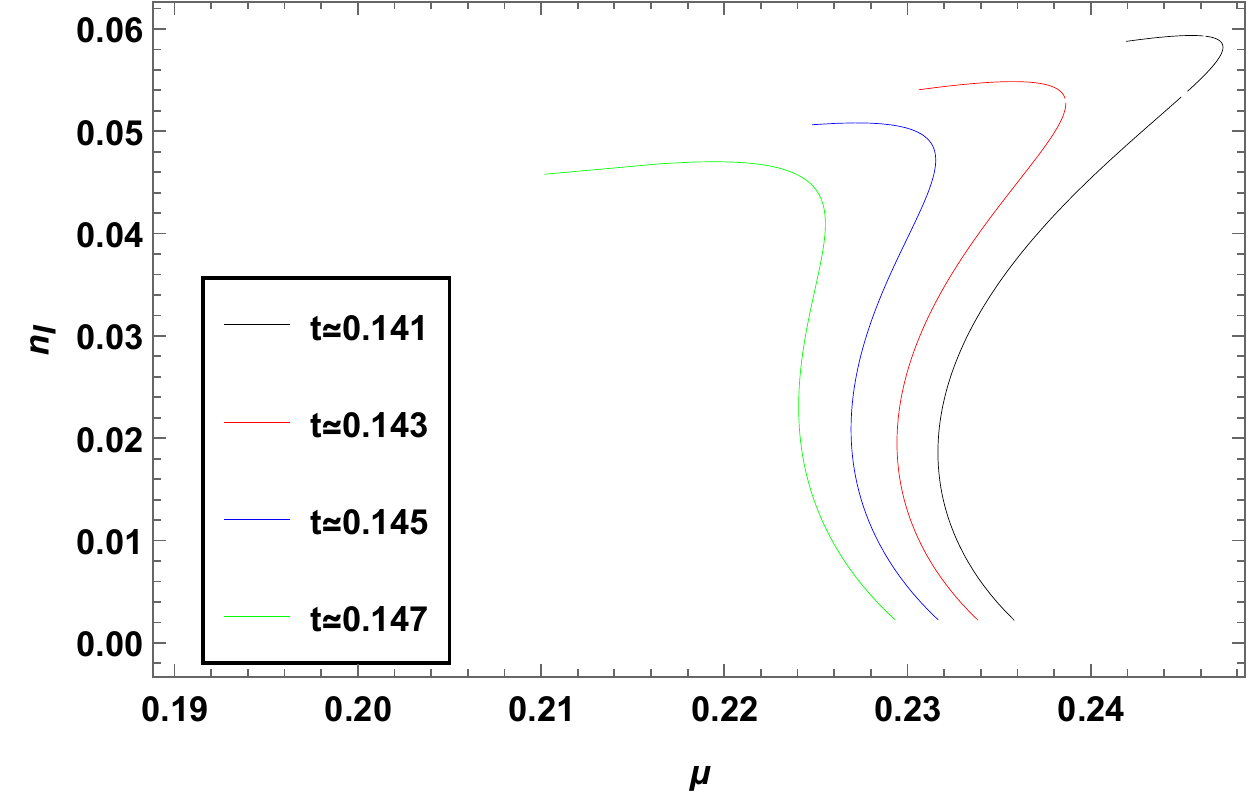}}
\caption{\label{nimu}Deconfined geometry at nonzero temperature, baryon density $n_{I}$ with respect to $\mu$ for amount of small and large temperature. Instanton width $\rho_{0}=4.3497$ and instanton deformation $\gamma_{0}= 3.7569$ and 't Hooft coupling $\lambda=15.061$, this data is when we consider the interaction term and lattice structure is fcc.}
\end{figure}

We use (\ref{26}) for deriving our results with or without the interaction between the instantons, except for calculating the thermal baryon mass.
\item{Baryon density}\\
Because of numeric approach for solving the main equations, we should check our results carefully. For this reason we calculate the number density of instantons or baryon density, $n_I$ from the following equation
\begin{equation}
n_{I}=-\frac{\partial \Omega_{baryon}}{\partial \mu},
\label{check}
\end{equation}
where $ \Omega $ is the baryonic free energy and $ \mu $ is chemical potential. We do this calculation one time to make sure that our results match together, in Fig. \ref{nimu}, we show results  for small  temperatures $t\simeq0.05,0.07, 0.08, 0.1 $ and large temperature $ t \simeq 0.141, 0.143, 0.145, 0.147$  and choice of parameters $\rho_{0}=4.3497$ and $\gamma_{0}=3.7569$,$\lambda=15.061$. Of course, this diagram can be drawn for other parameters, but, if we change the parameters $ \rho $ and$ \gamma $ the type of transition will be different, in some regions we will have a second-order phase transition. For example, at the temperature of $ t=0.1 $ and $ \rho=1.5 $ and $\gamma=2  $ the phase transition from the meson to the quark is second order, generally for $ \rho=1.5 $ in the area of $ 0.8 \leq \gamma \leq 3 $ at $ t=0.1 $ we have a second-order phase transition.\\
In the lower temperature at large chemical potential, we have a second turning point and then, we have unstable phase.
The features that our selected parameters have, by considering $ M_{KK}=\frac{3185MeV}{\pi} $ and $ \ell=1 $, is that with properties of nuclear mater have been fitted such as baryon mass, binding energy, saturation density and incompressibility. As it is clear in the left panel, there is a turning point where shows the first order baryon onset for all temperatures, the calculation that leads to this Fig. \ref{nimu} was first presented in Ref. \cite{Schmitt:2015} for $t=0$ and we recapitulate it for $t\neq 0$, as the figure shows, the geometry of the plot does not change when the temperature is taken into account. Only small amounts are changed. For the small chemical potential for temperatures $ t\leq 0.1 $, baryonic phase is preferred. But in the right panel, we plotted $n_{I}$ with respect to $\mu$ for large temperatures, for largest temperature, there is not turning point and probably we have second order phase transition. But these solutions are not stable and chirally symmetry phase is preferred.\\
\end{itemize}
\section{Results}
\subsection{Phase diagram}
In this section, we use equations for baryonic, quark and mesonic free energy and study the phase diagram of the system on the plane of temperature and chemical potential. We assume that the baryons have the size and also the dimensions of the baryon are temperature dependent, and the interaction between the baryons is considered.\\
At first, we shall explain the three phases:
\begin{itemize}
    \item Mesonic phase\\
    In this phase, $D8$ and $\bar{D}8$ branes connected at $u=u_{0}$ and $x_{4}^{\prime}(u_{0})=\infty$, also in this phase boundary condition is:
    \begin{equation}
        x_{4}(u_{0})=0,\ \ \ \ \ \ \ \ \ a_{0}^{\prime}(u_{0})=0,\ \ \ \ \ \ \ a_{0}(\infty)=\mu,
    \end{equation}
    and number density $n_{I}=0$, so the finite temperature result of free energy for mesonic case become:\\
    \begin{equation}\label{freemesonic}
        \Omega_{mesonic}=\int_{u_0}^\infty du \frac{{u^{13/2}}\sqrt{f_{T}(u)}}{\sqrt{u^8f_{T}(u)-u_{0}^{8}f_{T}(u_{0})}}.
    \end{equation}
    \item Quark phase\\
    In chirally restored phase, flavor branes are not connected, in this phase we have $x_{4}^{\prime}=0$ and the boundary conditions:
   \begin{equation}
         a_{0}(u_{T})=0,\ \ \ \ \ \ \ a_{0}(\infty)=\mu,
    \end{equation}
    and free energy becomes,
    \begin{equation}\label{freequark}
        \Omega_{quark}=\int_{u_T}^\infty du \frac{u^5}{\sqrt{u^5+n_{I}^{2}}}.
    \end{equation}
    In the above relation, $n_{I}$ is as a function of $\mu$ and $T$, so one can obtain,
    \begin{equation}\label{muquark}
        0=\mu-\frac{n_{I}^{2/5}\Gamma(\frac{3}{10})\Gamma(\frac{5}{5})}{\sqrt{\pi}}+u_{T}{}_2 F_1\left[\frac{1}{5},\frac{1}{2},\frac{6}{5},-\frac{u_T^5}{n_I^2}\right],
    \end{equation}
    but also, you can see detail of mesonic and quark matter phases in appendix B of Ref. \cite{Schmitt:2015}.

    \item {Baryonic phase}\\
    We discussed this phase in the last section and we introduced \eqref{free} for free energy of baryonic phase.
\end{itemize}

\begin{figure}
\centering
\includegraphics[width=.6\textwidth]{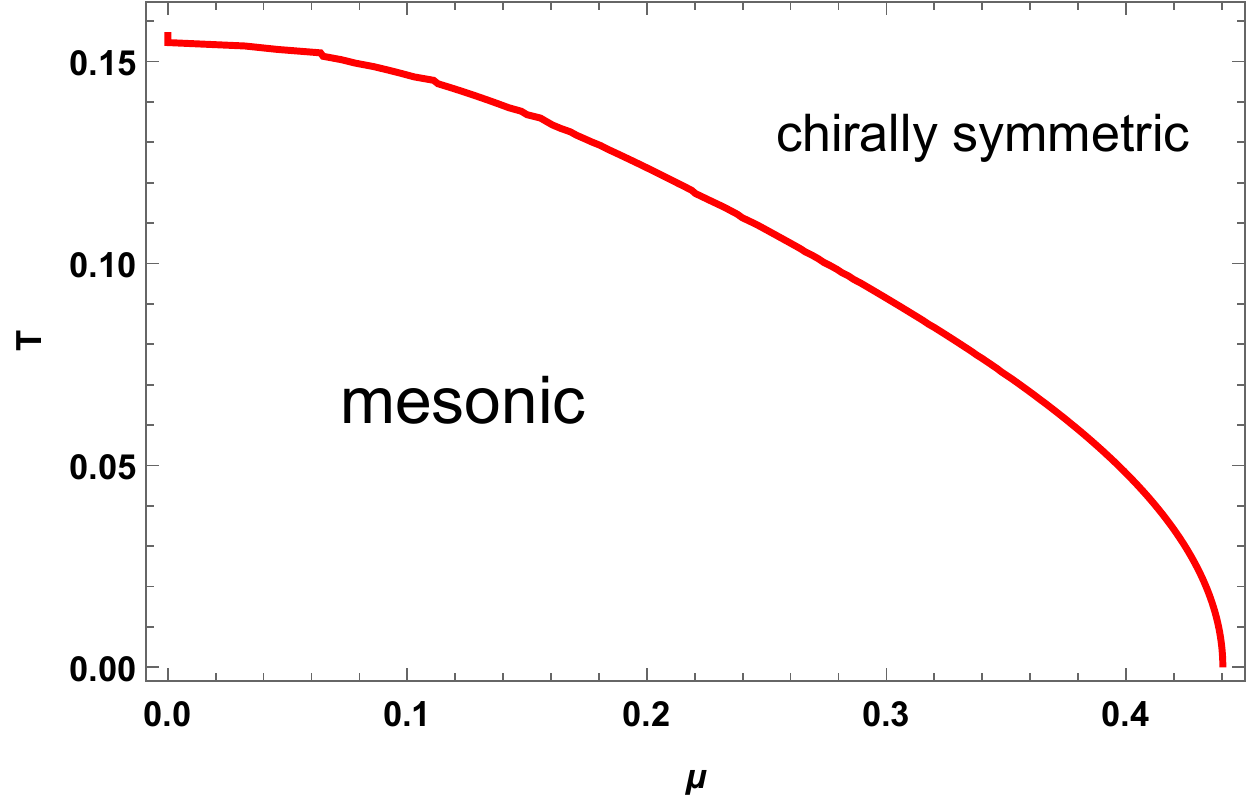}
\caption{\label{phase2}Phase diagram that separate confined and deconfined phase in the plane of dimensionless temperature $T$ and chemical potential $\mu$. }
\end{figure}
\begin{figure}
\centering
\hbox{\includegraphics[width=.5\textwidth]{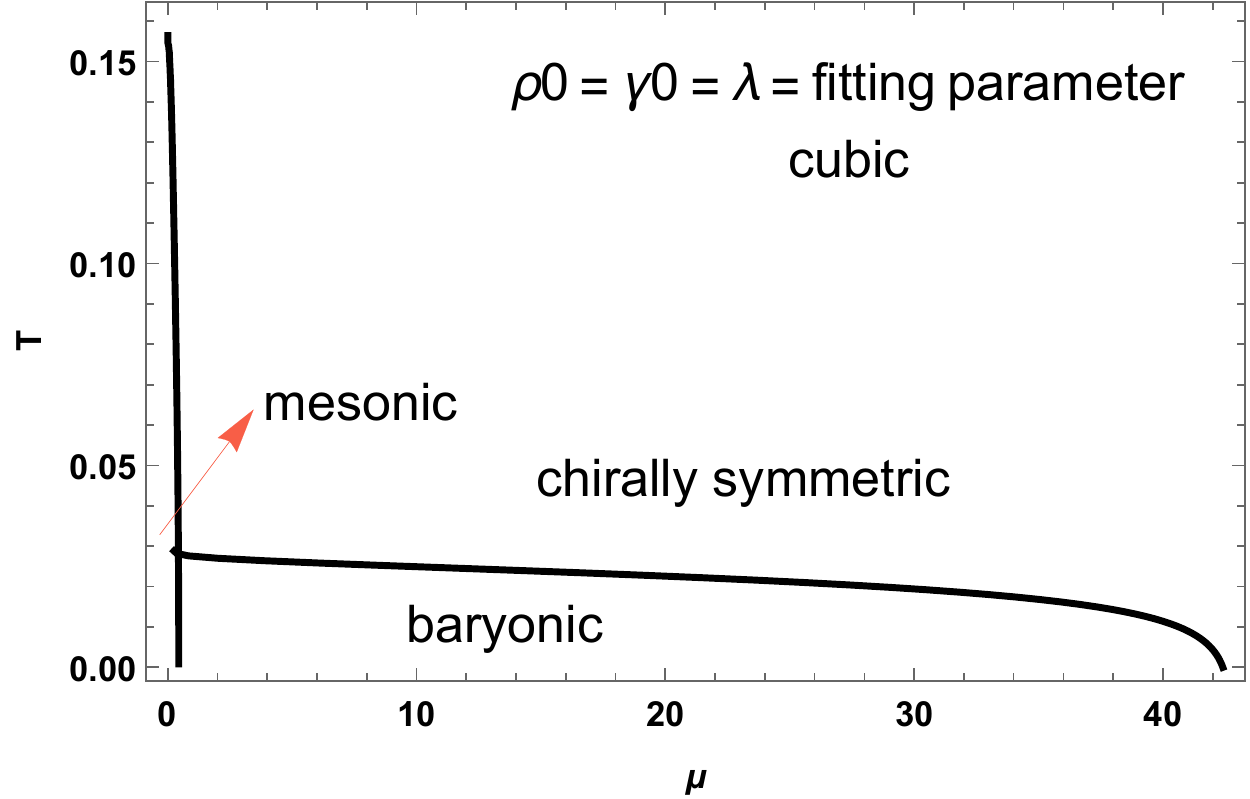}\includegraphics[width=.5\textwidth]{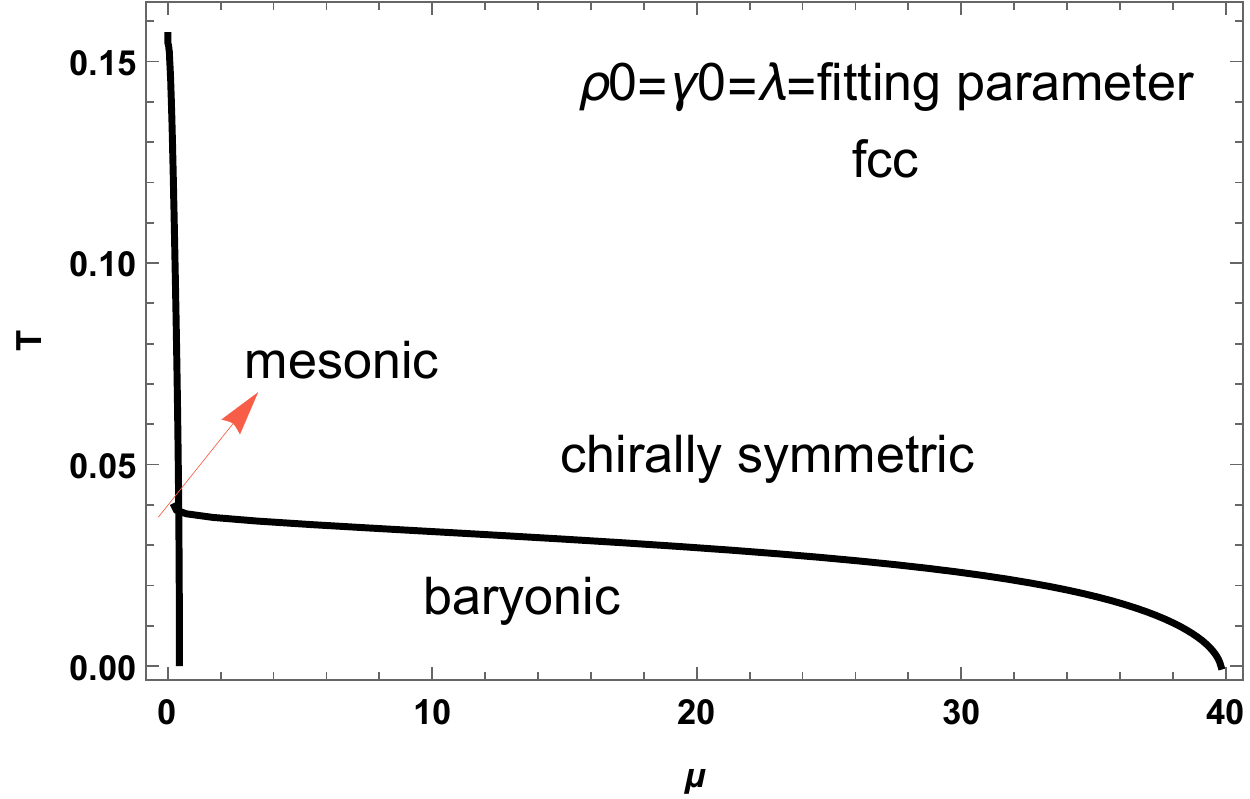}}
\caption{\label{phase3} Phase diagram for different lattice structures cubic and $fcc$. $\lambda$, $\rho$ and $\gamma$ are fitting parameters. It is in the deconfined geometry when considering baryon and Interaction between them.}
\end{figure}

First, we explain how  can calculate the phase transition lines in Fig. \ref{phase2} and Fig. \ref{phase3}. We have solved equations \eqref{minimOm}  and also the condition that they have the same free energies (which means the transition between two phases happens when $\Omega_{baryon}=\Omega_{quark}$ for baryon-quark phase transition line  and $\Omega_{meson}=\Omega_{quark}$ for the meson-quark phase transition line).
We have to calculate the initial value at temperature $t=0$ also you can see the relation between capital letter $T$ and small letter $t$ in table 1. That is, find the free energy of quark, baryon and meson, and then by equating the free energy, find the starting points at temperature $t=0$.\\
In Fig. \ref{phase2}, we show plot of the phase diagram that separates the mesonic phase from the quark, and the red line is the  phase transition  line from mesonic to quark. As the same as  \cite{Horigome:2007} we work in chiral limit that means  quark mass is  zero and $\mu(t=0)=0.44$ is first order phase transition from meson to quark.
However, in Ref. \cite{Schmitt:2020} they have investigated a case where  quark mass is non-zero. In this case the value of the phase transition is dependent to a parameter that related to quark mass.\\
In Fig. \ref{phase3} we study the phase diagram in the presence of Baryons and using \eqref{shapeT}, i.e the case where the parameters $ \rho $ and $ \gamma $ are temperature dependent. One should notice that these parameters were also dependent on the t'Hooft coupling constant $ \lambda $. In Fig. \ref{phase3}, we show the phase diagram for different lattice structure, we fix $\rho_0=4.3497$, $\lambda=15.061$, $\gamma_{0}=3.7569$ and $M_{KK}=3185/\pi$. These parameters are fitted to saturation density, vacuum mass of the nucleon, binding energy and incompressibility. By changing lattice structure, The chemical potential also changes significantly where the baryonic matter changes to the quark phase  ($\mu_c$), which is in agreement with the result of Ref. \cite{Preis:2016gvf}. We  find phase transition value by calculating free energy for baryonic and quark cases, when they are equal a first order phase transition happens at this value $\mu$.\\Based on phase diagram, we will find that the least phase transition value at $ t=0 $ is $\mu(t=0)\varpropto40 $ for fcc lattice structure, according to table 1, Kaluza-Klein mass, critical chemical potential is $\mu(t=0)\varpropto 40000 $ MeV. These values are not within the range of the chemical potential of the neutron star, $300<\mu<500  $ \cite{Schmitt:2010} so if we want to be included in the neutron star range should be adjusted so, fortunately, these values depend on the model parameters according to the phase diagram  can be changed by $\rho_{0}$ and $\gamma_{0}$ or $\lambda$
and lattice structure found. Of course, these changes in the model have caused that, contrary to the diagram in Fig. \ref{nimu} of the paper \cite{Schmitt:2015}, in the large  chemicals potential  and low temperature, chiral symmetry has been restored, in which case the physical principles have been observed.\\
Table 2 is the chiral phase transitions value ($ \mu_{ph} $) for both fixed temperature $t=0.058$ and different 't Hooft coupling ($\lambda$). According the AdS/QCD dictionary we have $ \lambda=g^{2}N_{c} $, so one find that the chiral phase transition is expected to occur at moderate, not asymptotically large density. It means that the model may be invalid in the large chemical potential.\\

\begin{table}\label{table2}
\centering
\begin{tabular}{ | c| c| }
\hline
\hline
  $ \lambda $ & $ \mu_{ph} $ \\
\hline
 10 & 46.6 \\
 \hline
 18 & 39.7  \\
 \hline
 40 & 36.3  \\
 \hline
 50 & 35.2 \\
\hline
\end{tabular}
\caption{We show chiral phase transition value for different lambada in interaction case.}
\end{table}
\subsection{Connecting nuclear matter with quark matter at finite temperature}
\begin{figure}
\centering
\hbox{\includegraphics[width=.5\textwidth]{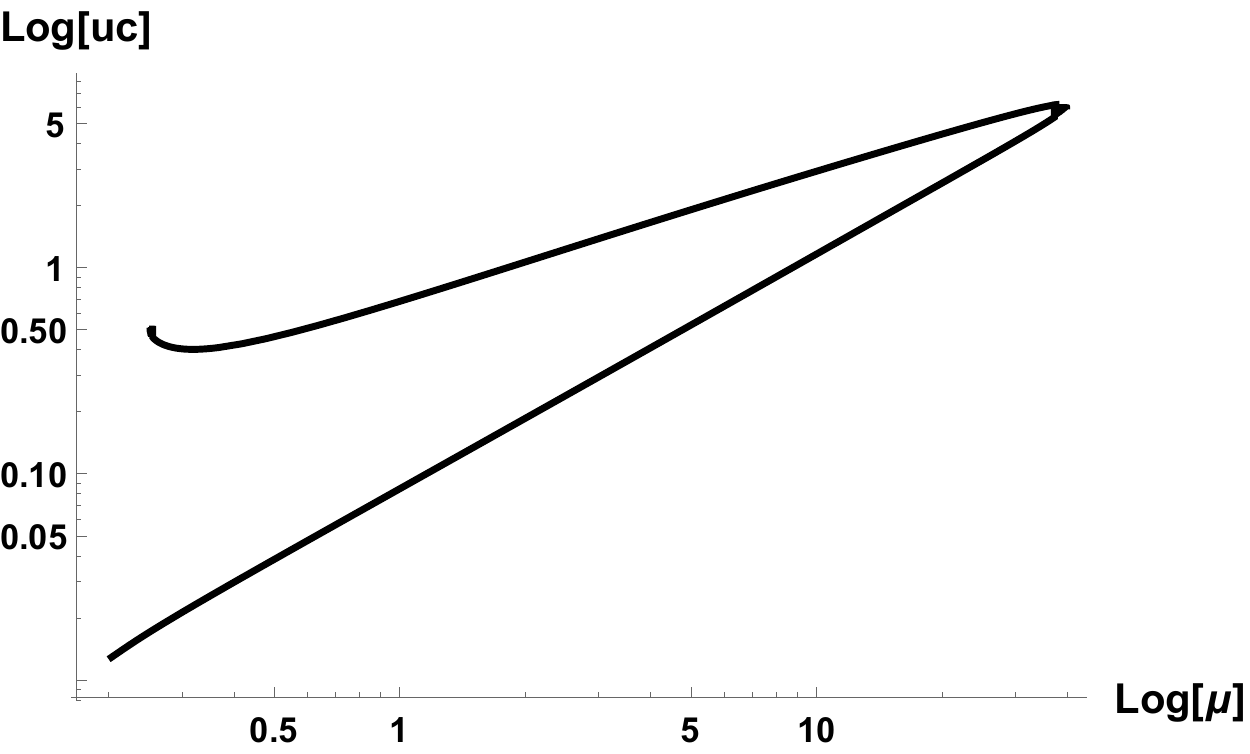}\includegraphics[width=.5\textwidth]{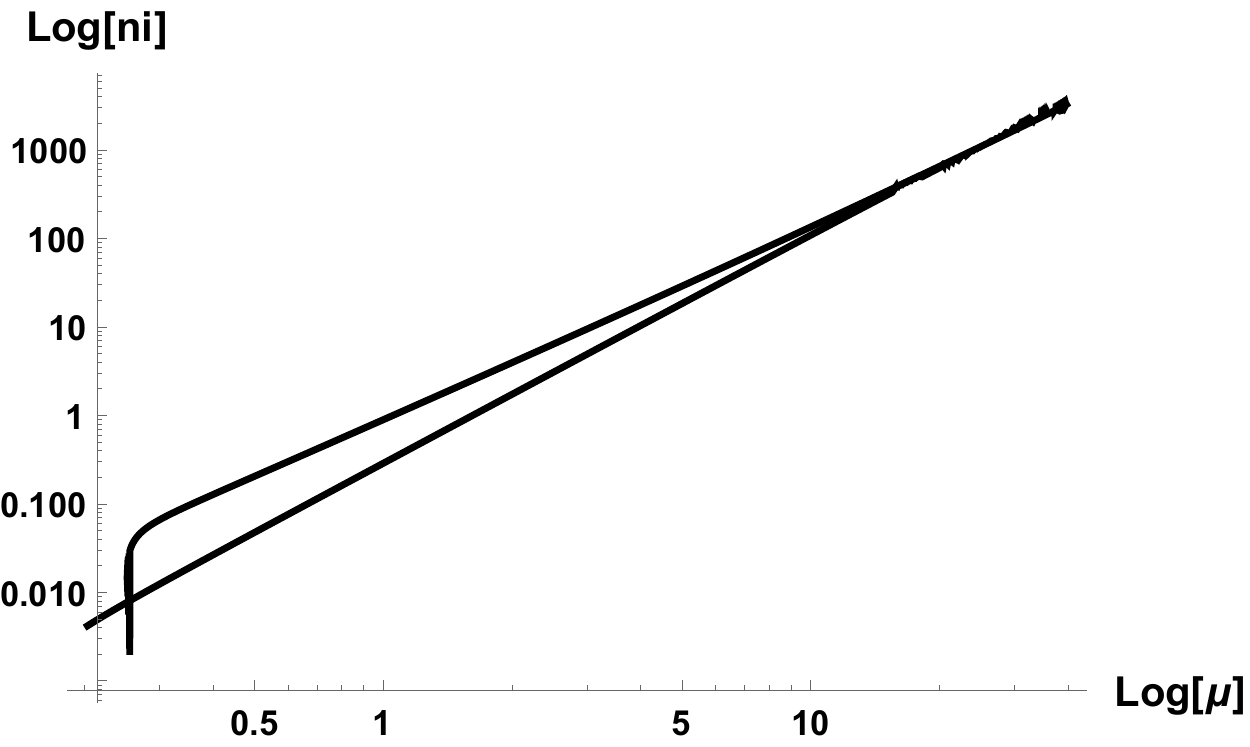}}
\caption{\label{continuity}Location of the tip of the connected flavor branes $u_{c}$ (left panel) dimensionless baryon density $n_{I}$ (right panel) as a function of the dimensionless chemical potential $\mu$ at nonzero temperature.  }
\end{figure}

\begin{figure}
\centering
\vspace{0.2cm}
\hbox{\includegraphics[width=.5\textwidth]{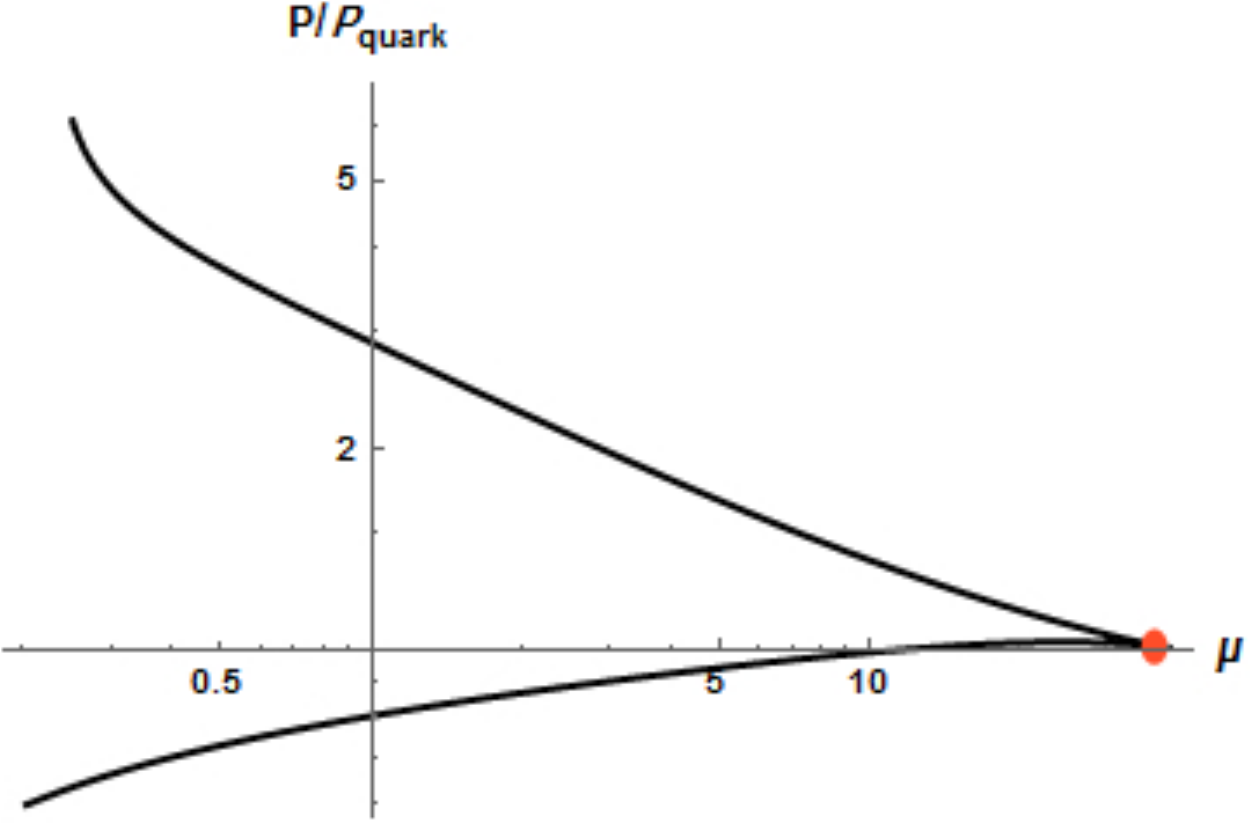} \includegraphics[width=.5\textwidth]{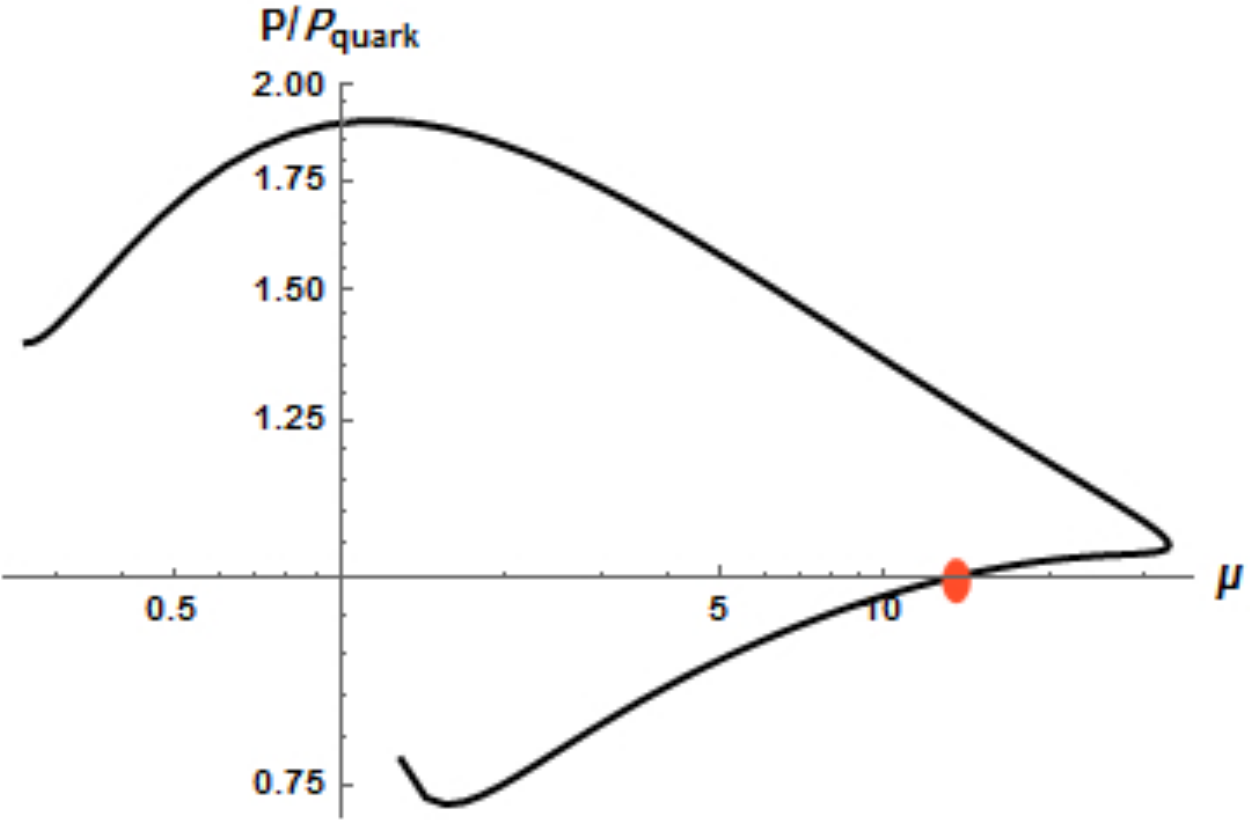}}
\hbox{\includegraphics[width=.5\textwidth]{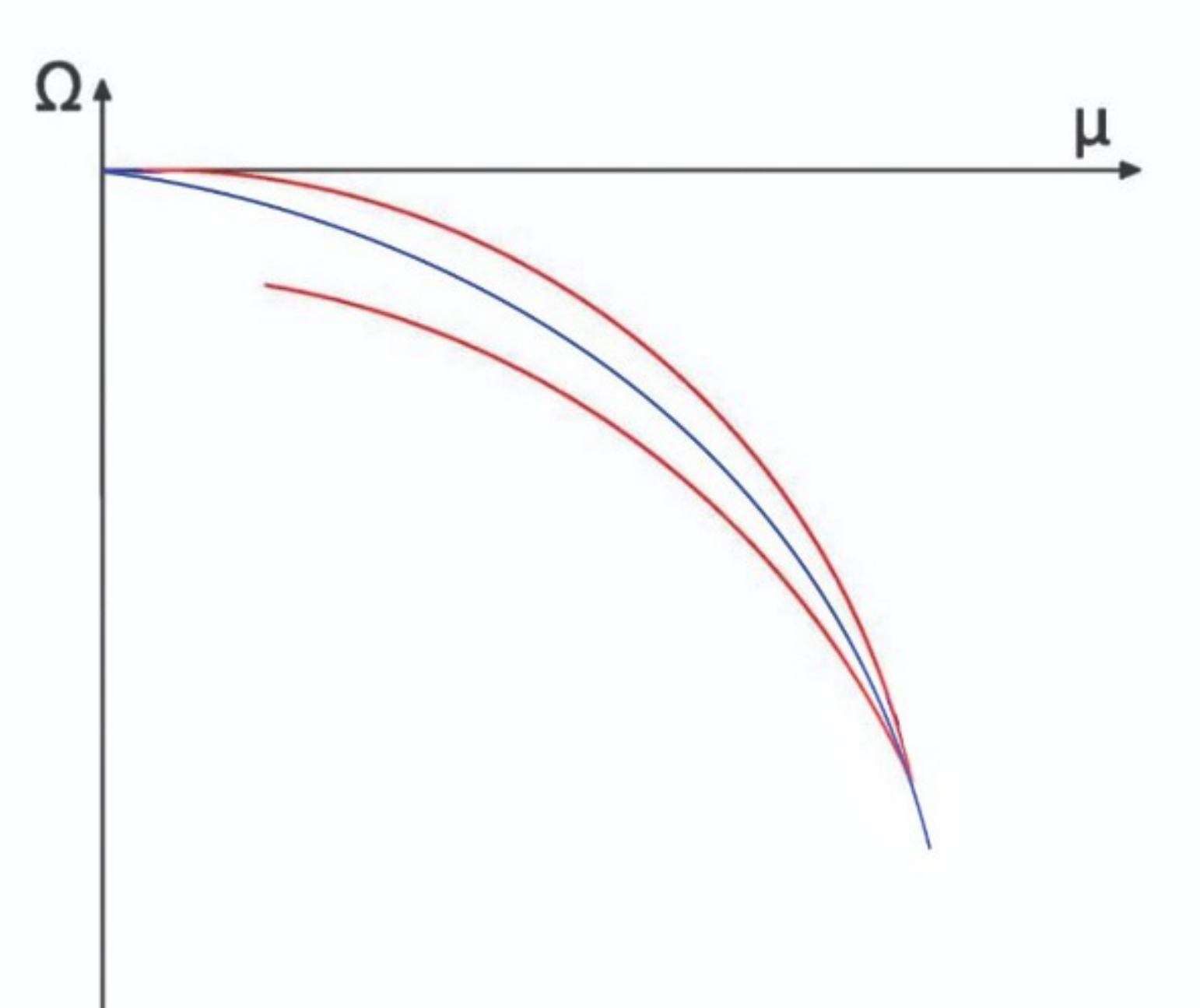} \includegraphics[width=.55\textwidth]{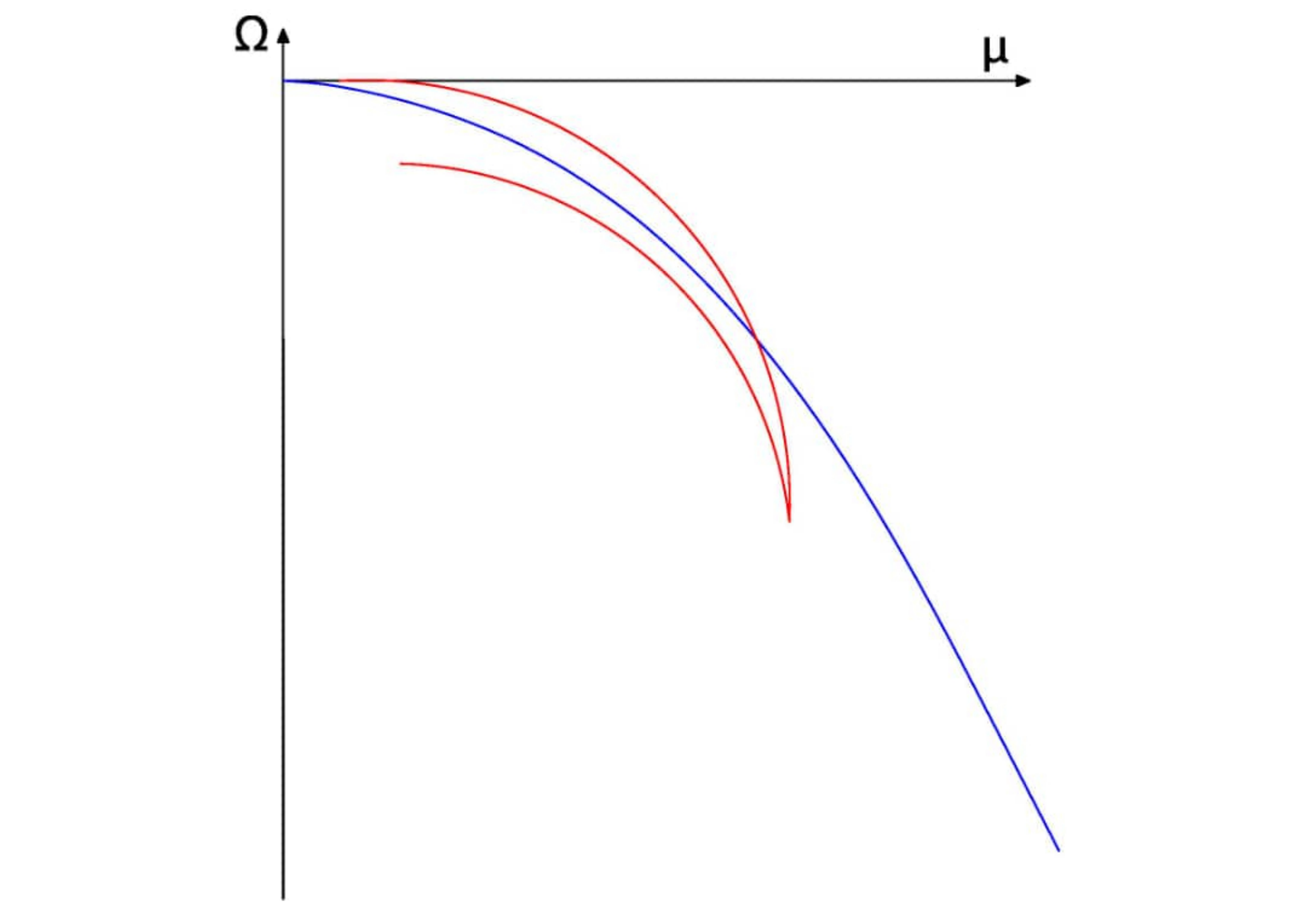}}
\hbox{\includegraphics[width=.5\textwidth]{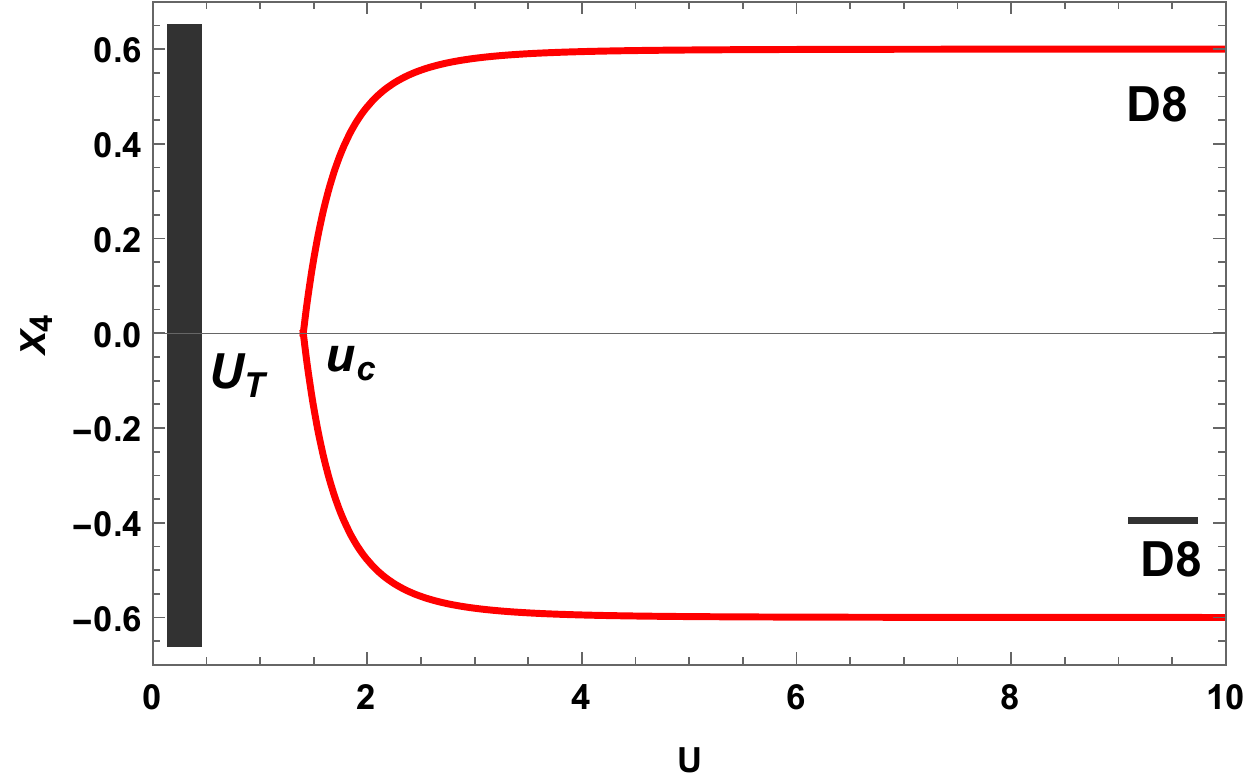} \includegraphics[width=.5\textwidth]{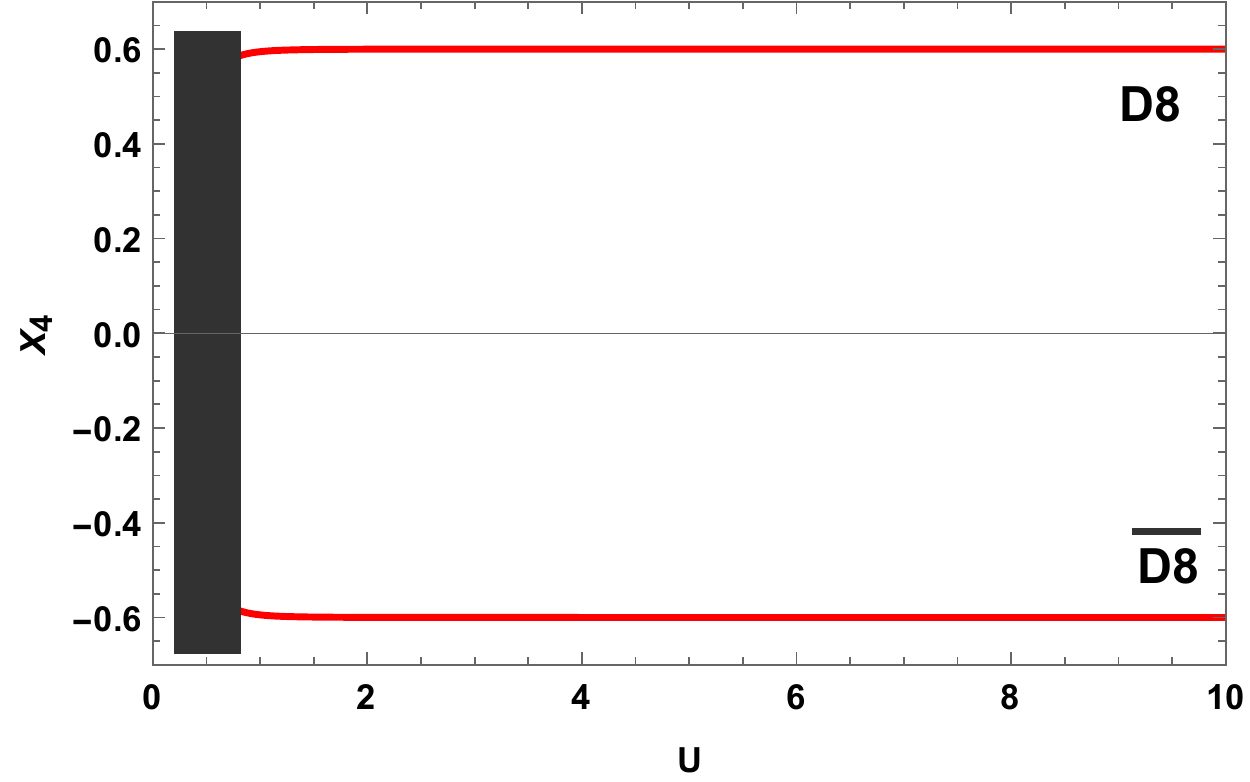}}
\caption{\label{continuity1}First row: The pressure of the baryonic phase normalized to the pressure of the quark matter phase for non-zero temperature. First column is for $t\approx0.024$, and second column is for $t\approx0.086$. Also, the red circle is critical chemical potential from baryonic to quark. Second row: the cartoon plot of free energy as a function of $\mu$ for the baryonic phase.  $\rho_{0}=4.3497,\gamma_{0}=3.7569,\lambda=15.061 $ and lattice structure is fcc. Third row: geometry of $D8$ and $\Bar{D}8$ in the subspace $(x_{4},u)$, black region indicate the location of the horizon. }
\end{figure}

In Ref. \cite{bitaghsir:2018kfa}, the relationship between nuclear matter and quark matter was investigated at zero temperature. In this section, we examine this state in the presence of temperature.\\
Three curves $[d(\mu), u_{c}(\mu), n_{I}(\mu)]$ have described the solutions of the free energy and we have a route that would have existed from the baryonic phase into the chirally symmetry phase when chemical potential is not monotonic.\\
In Fig. \ref{continuity}, we represent the all numerical solutions. We have plotted $n_{I}$ (right panel) and $u_{c}$ (left panel). We haven't plotted $k$ because of this variable have a qualitatively identical manner like $u_{c}$. We have used a double-logarithmic plots to make full important features of the solutions visible. We see that large chemical potential, $n_{I}$ and $u_{c}$ become multi-valued in the baryonic phase and particular interpretation for this behavior is a first order baryon onset.\\
We had similar plots in Ref. \cite{bitaghsir:2018kfa} except that there were solutions to the small chemical potential  in the plots, and it was because of the analytical solutions we found at these points, but when the temperature is taken into account, these results  is not at small chemical potential. Of course, these solutions will be unstable due to the fact that the mesonic phase has a smaller free energy then the mesonic phase is preferred.\\
In the First row of Fig. \ref{continuity1}, we have the pressure $p=-\Omega$ for the baryonic phase (black curve) that divided by the pressure of quark phase for non-zero temperature $t\approx0.024$ and $t\approx0.086$. After baryon onset with considering interaction case, we can see turning point.\\ The preferred state for a given chemical potential is the one with the largest value of $P/p_{quark}$. Red dot in the First row of Fig. \ref{continuity1} have been shown the transition between baryonic to quark phase.\\
\begin{figure}
\centering
\hbox{\includegraphics[width=.5\textwidth]{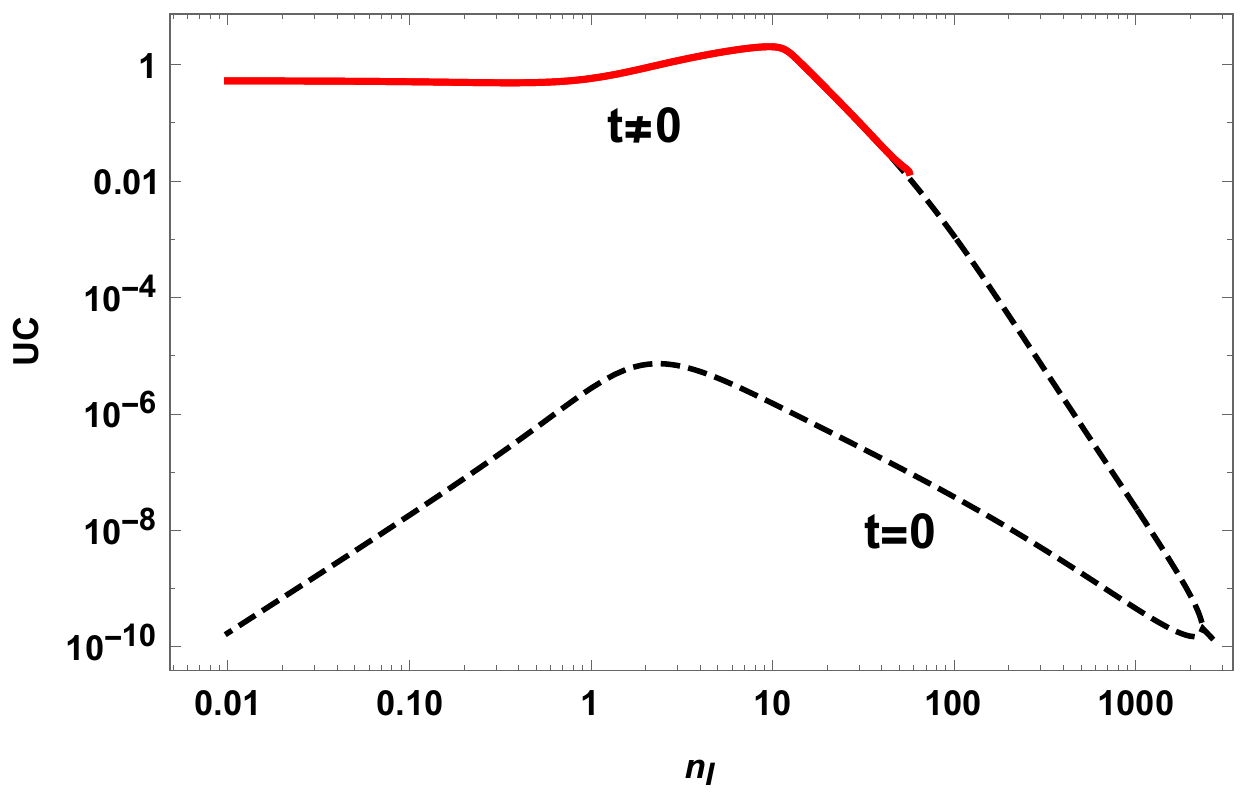}\includegraphics[width=.5\textwidth]{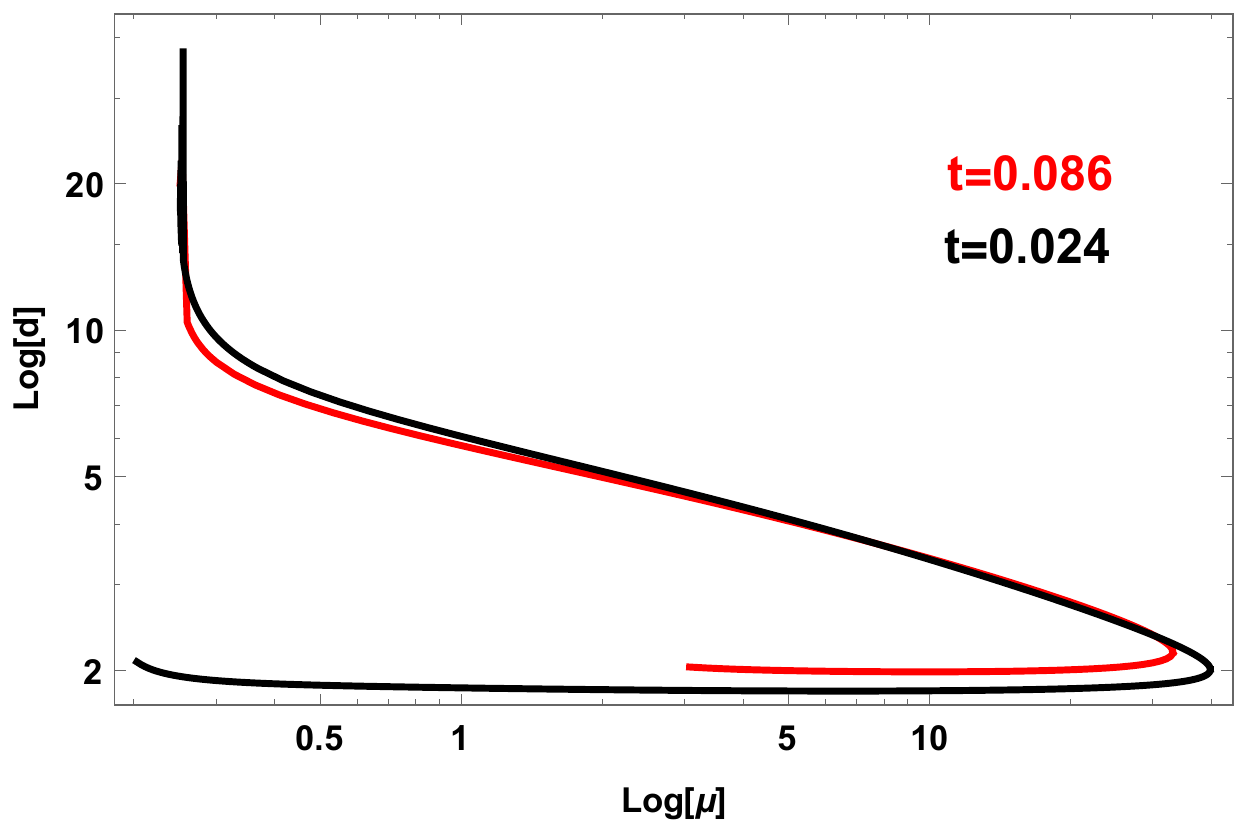}}
\includegraphics[width=.5\textwidth]{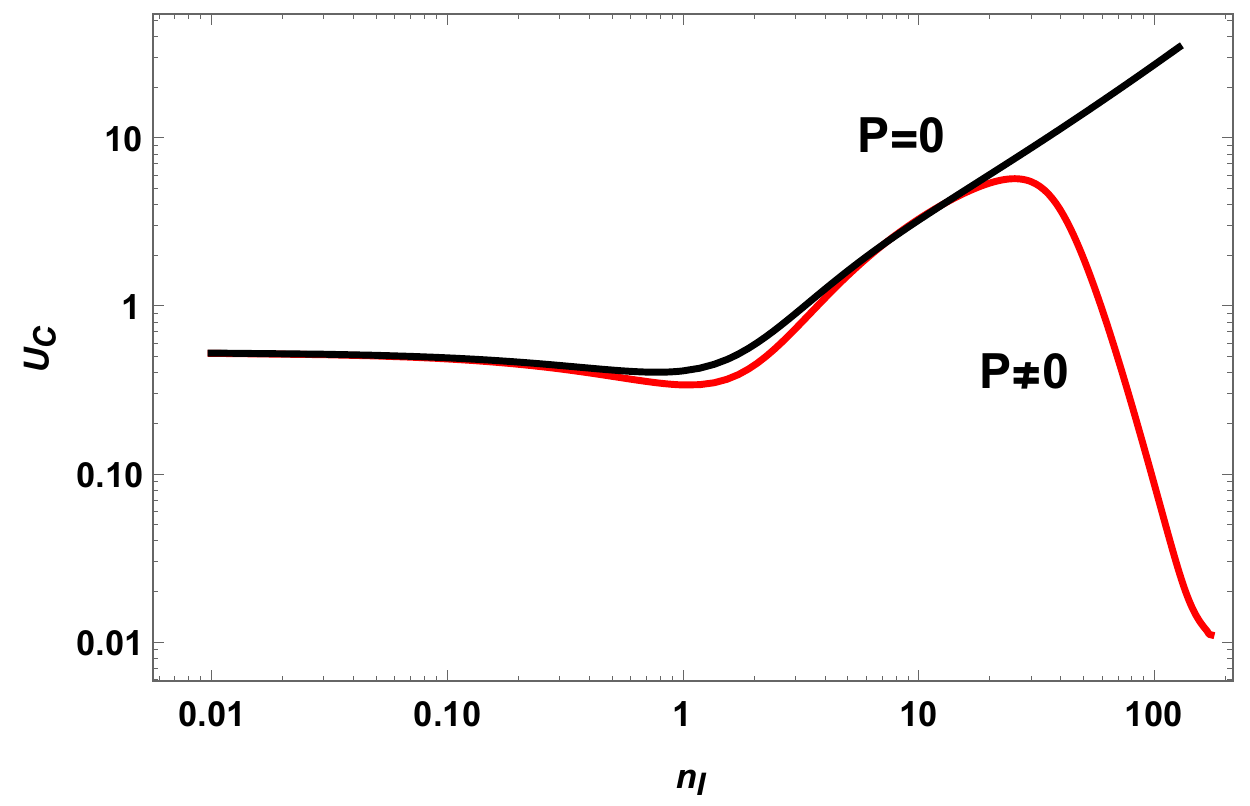}
\caption{\label{continuity2} First row: left panel: location of the tip of the connected flavor branes $u_{c}$ as a function of dimensionless baryon density $n_{I}$ for red curve is $t\neq 0$, and dashed black curve is $t=0$. Right panel: instanton distance over instanton width $d$ as a function of $\mu$ for baryonic phase for $t=0.024,0.086$ temperature.
Second row: location of the tip of the connected flavor branes $u_{c}$ as a function of dimensionless baryon density $n_{I}$ for interaction case ($P\neq 0$, red curve) and non interaction case ($P=0$, black curve).}
\end{figure}
In the second row of Fig. \ref{continuity1}, we show cartoon plots of $\Omega$ with respect to $\mu$ that red curve related to the solutions of equation \eqref{free} and show baryonic state. But, blue curve is quark phase. Critical chemical potential is point that two curves cross each other and have been existed transition between baryonic to quark phase. In $t=0.024$, chemical potential for first order phase transition is around $39/\ell^2$ but in $t=0.086$, this point is unphysical because of upper branch in baryonic matter is unstable because, it has the largest free energy of all solutions. By consideration of  nucleon mass, they are placed at nonzero chemical potential, so upper branch in second row of Fig.  \ref{continuity1} should not be confused with ordinary nuclear matter where the density is reduced as we approach the origin. From a certain temperature onwards for chosen parameter, the transition point from baryonic state to  quark becomes  unphysical because in real QCD at high temperatures we have only quark matter.\\
The disfavoring phase, at small chemical potential is quark phase compared to baryonic phase, therefore, in the second row of the Fig. \ref{continuity1}, at first, red curve is favored compared to the blue curve. Also, we can see the quark matter solution in all path and infinitesimally small chemical potential because of we work in the chiral limit.\\ Considering all solutions of free energy, all branches from nuclear matter phase to quark matter phase are continuously connected. It should be noted that solutions are not necessarily stable.\\
In the third row of this figure, the geometry of the branes $D8$ and $\Bar{D}8$ have been shown for two different temperature modes and exactly at the transition point (ie the red dot in the first row). The black border has shown black hole in the bulk  that this value increases and decreases for different temperatures, and in $t=0,024$, $D8$ and $\Bar{D}8$ branes are connected and $u_{c}$ is a maximum value , so this point does not reach the black hole, but in $t=0.086$ amount of $u_{T}$ increases until $u_{T}=u_{c}$ these branes are melted in the black hole, so the chiral symmetry is restored and we have a unstable baryonic phase.\\
In all figures, the axes labels have to be include suitable powers of $\ell$, i.e for $n_{I}$ is $n_{I}\ell^{5}$, $\mu$ for $\mu\ell^2$, $u_{c}$ for $u_{c}\ell^{2}$ and $\Omega$ for $\Omega\ell^7$ but overlap parameter $d$ doesn't scale with $\ell$.\\
The parameter set is for $\rho_{0}=4.3497, \gamma_{0}=3.7569, \lambda=15.061 $ that this parameter set satisfy physical constraints Ref. \cite{bitaghsir:2018kfa}, lattice structure for both row are face centered cubic crystal that means we have $p=12$ and $r=\sqrt{2}$.\\
The right panel of  Fig. \ref{continuity2} shows instanton distance over instanton width $d$ as a function of $\mu$ for baryonic phase. The black curve is related to $t=0.086$ and red one is related to $t=0.024$, and also left panel is the location of the tip of connected flavor branes $u_{c}$ as a function of dimensionless baryon density $n_{I}$ for $t=0$ and $t\neq 0$. For interaction case $p\neq 0$, and for non-interacting case we have $p=0$. The results show that we have turning point for $p\neq 0$ and but for non-interacting ($p=0$) case finished in the large $n_{I}$ and numerics get increasingly complex and we do not see any turning point.\\
According to the dimensions of instanton, we expand them then we have
$$ \rho\varpropto\frac{1-\frac{9 u_{T}^{3}}{16u_{c}^{3}}}{1-\frac{1u_{T}^3}{32u_{c}^3}}\rho_{0},$$
and
$$ \gamma\varpropto\frac{3}{2}\gamma_{0}(1-\frac{5u_{T}^{3}}{16u_{c}^{3}}).$$
As it is known, instanton shape $ \rho $ and $ \gamma $ have explicit dependence on temperature with $ u_{T} $ parameter and they have implicit dependence through the dynamical value of $ u_{c} $. Considering the temperature, the amount of $ \rho $ and $ \gamma $ decreases and as a result, the dimensions of instanton under the space of $(x_{1},x_{2}) $ and $ (z,x_{1}) $ changes and because the temperature is always greater than zero, the dimensions of the instanton have a non-zero width in the holographic direction and also a limited in the spatial direction.

\subsection{Speed of sound}
The speed of sound $ C_{s} $ is defined via the following equation,
\begin{equation}
C_{s}^{2}=\frac{\partial P}{\partial \varepsilon}.
\end{equation}
Using $ P=-\Omega $ and $ \varepsilon=\Omega +\mu n+st $, we plot $C_s^2$ for different values of temperature. Also, we change temperature by changing $ u_{t}$, we know that consider the baryonic phase and the parameters $ \rho_{0} $
and $ \gamma_{0} $ are chosen in a way that is fitted by the nucleon mass
for parameters.
\begin{figure}
\centering
\includegraphics[width=.5\textwidth]{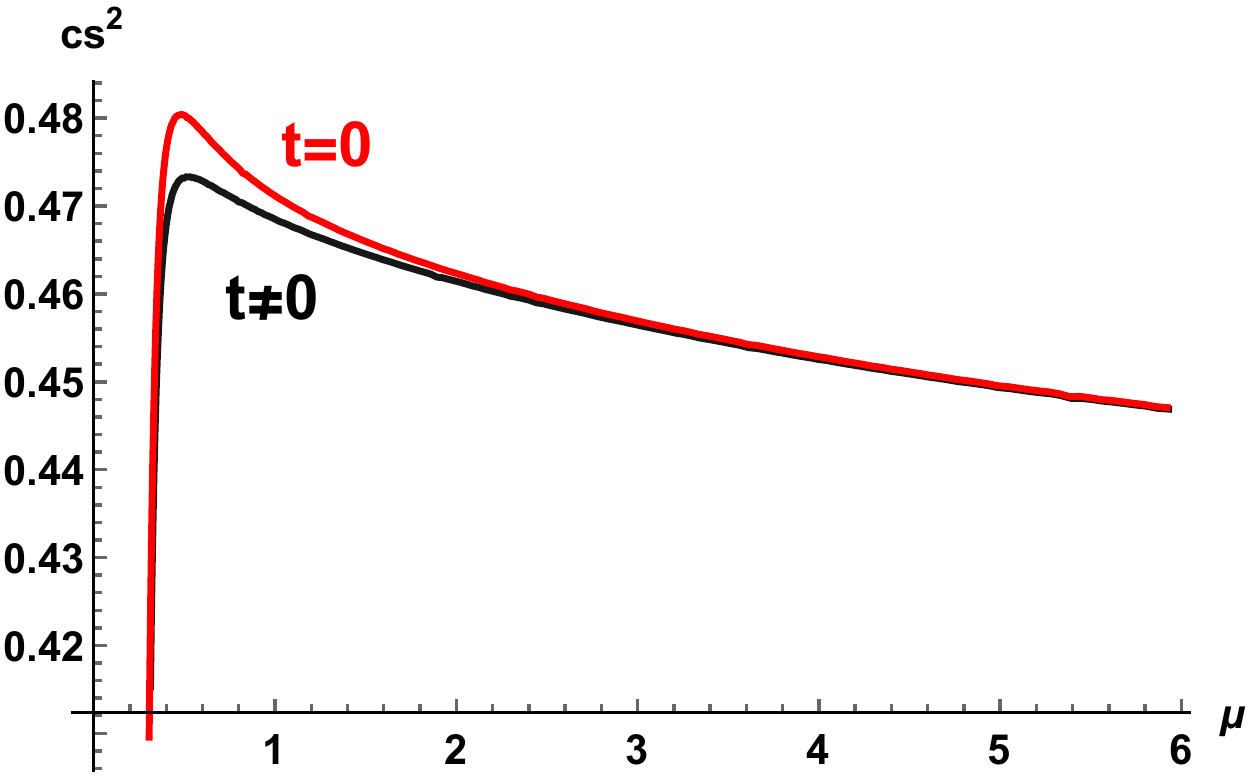}
\caption{\label{sT}Speed of sound as a function of chemical potential for different temperature $t=0$ (red line) and $t\neq 0$ (black line), we fixed $\rho_{0}\simeq 2.76$ and $\gamma_{0}\simeq 4 $.}
\end{figure}
Fig. \ref{sT} is a plot that shows the sound speed of the baryonic matter with respect to chemical potential. The red line of Fig. \ref{sT} dawn for the case where the temperature is equal to zero and the black line is represents the case where the temperature is non-zero. We assume that in this figure, the parameters of the model are $\rho_{0}\simeq 2.76$ and $\gamma_{0}=4$ and $\lambda=15.061$. For the both cases that are considered, the minimum temperature value is greater than zero, ie., $\simeq 0.4$. We know from causality that the speed of sound must always have an upper limit of $C_{s}^{2}<1$ and also shows us the thermodynamic stability that lower limit is  $C_{s}^{2}>0$ from Ref. \cite{Cherman:2009gb} it has been seen that the conformal limit gives us the value of $C_{s}^{2}=\frac{1}{3}$ as can be seen from the plot, this limit is not observed in the baryonic matter and we have the value of approximately $\simeq 0.45$. However, it should be noticed that the quark phase is not taken into account. One of the features of this plot is that their minimum value is always greater than zero, which shows thermodynamic stability. As can be seen from the figure, in both temperatures is $0<C_{s}^{2}<1$, which indicates physical range. The speed of sound was calculated for the other parameters of the model, ie,  $\rho_{0}\simeq 6.49$, $\gamma_{0}=3$, $\lambda=15.061$ and similar results were obtained. These parameters are matched based on the mass of the baryon.

\section{Summary}
In this paper, we use the Sakai-Sugimoto holographic model. In this model, baryons are described by the instantons in bulk. Shape of instantons are defined by parameters $\rho$ and $\frac{\rho}{\gamma}$, $\rho$ is the width in the holographic directions and  $\frac{\rho}{\gamma}$ is spatial width. These parameters are expressed to upset the symmetry $SO(4)$. In this paper, what has been done is that the parameters $\rho$ and $\gamma$ are considered temperature-dependent. The temperature dependence is determined by the relations \eqref{shapeT}. This dependence causes let sentences be added to the entropy quantity, and let the principal equations give us other solutions under the influence of this dependence. According to these solutions, we study the thermodynamics of some properties such as the phase diagram, the quantity of the connection between the baryon and quark phase, and the speed of sound. It is observed that the effect of temperature on the parameters causes a change in the critical chemical potential from baryonic to quark state. It slightly changes the speed of sound in small chemical potentials. At some point, the speed of sound is not dependent on the chemical potential. Also, consider the temperature, there is still a connection between the baryon phase and the quark. In the references section, we also mentioned that the mass of baryons at different temperatures has different values. Indeed, we don't check the chiral phase transition in the small 't Hooft coupling, so it will be interesting issue for the future work.

\section*{Acknowledgments}
We would like to thank Andreas Schmitt and K. Bitaghsir Fadafan for valuable comments and discussion.
\appendix
\section{ Interaction profile of parameter}
We define parameters $h_{1}$, $S_{1}$, $h_{2}$, $S_{2}$, $a$ and $b$ of the equation (2.9) as follows,
\begin{align*}
h_1&\equiv 2\rho^{10}d^6(4d^4+7d^2-2)^2+2\rho^{8}z^2d^2(4d^2-1)^2(4d^6+12d^4+7d^2-2)  \nonumber\\
&+2\rho^6z^4(96d^{10}+116d^8-91d^6-14d^4+11d^2-1)+4\rho^4z^6d^2(26d^6-8d^4-23d^2+5)\nonumber\\
&+8\rho^2z^8d^4(2d^2-3)+4z^{10}d^4 \nonumber \\
h_2 &\equiv \rho^8d^4(4d^4+7d^2-2)^2+2\rho^6z^2d^2(32d^8+76d^6+15d^4-17d^2+2)\nonumber\\
&+\rho^4z^4(92d^8+144d^6+5d^4-18d^2+2)+8\rho^2z^6d^2(9d^4+8d^2-2)+28z^8d^4\, ,
\end{align*}
and
\begin{subequations}\label{SSab}
\bea
a&\equiv& z^2-\rho^2(d^2-1) \, , \qquad b \equiv \rho^2[4d^2(\rho^2+z^2)-\rho^2] \, , \\[2ex]
{\cal S}_1 &\equiv& \sqrt{\frac{-a+\sqrt{a^2+b}}{b}} \, , \qquad
{\cal S}_2 \equiv \sqrt{a+\sqrt{a^2+b}} \, .
\eea
\end{subequations}
\section{ Phase diagram for Non-interaction case}
\begin{figure}
\centering
\hbox{\includegraphics[width=.5\textwidth]{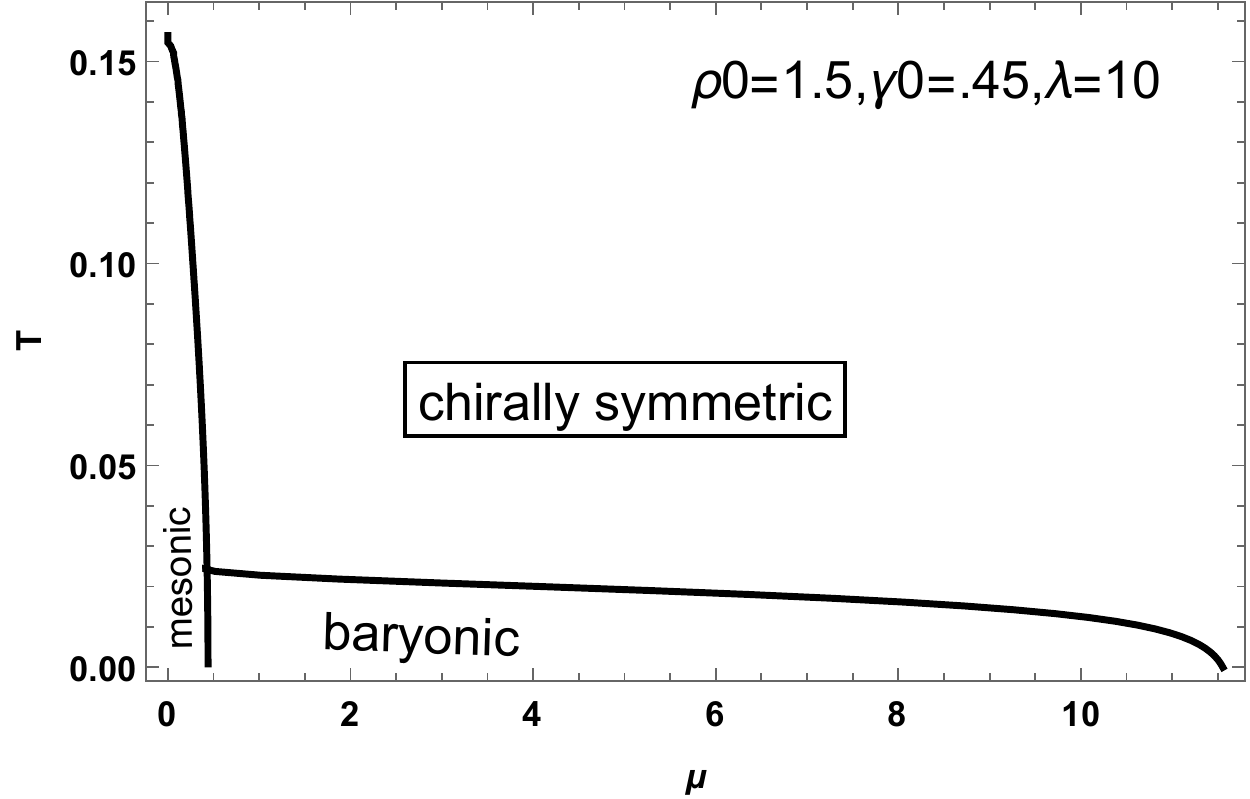}\includegraphics[width=.5\textwidth]{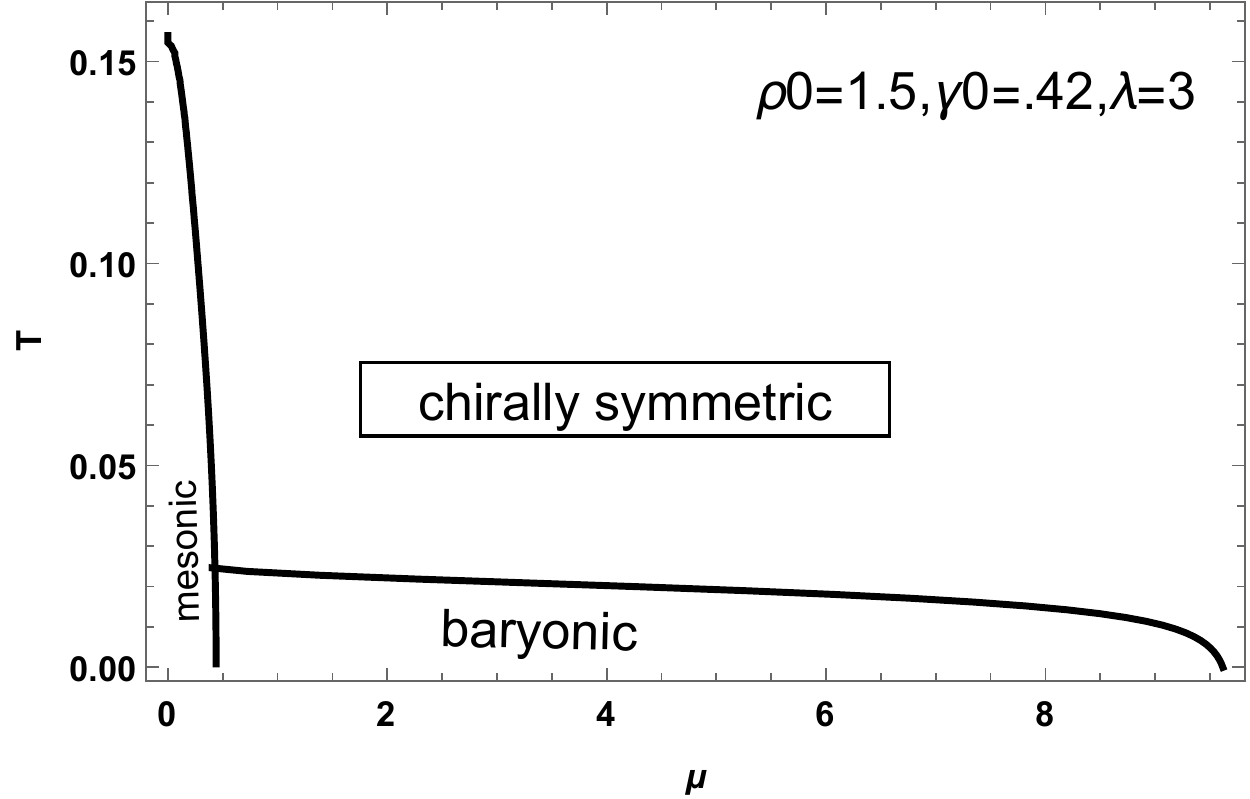}}
\caption{\label{phase1} Phase diagram for different parameters and in the deconfined geometry with baryon and non-interaction case.}
\end{figure}

We study the phase diagram in the presence of baryons in this case where the parameters $\rho$ and $\gamma$ are temperature dependent. One should notice that these parameters also dependent on the t'Hooft coupling constant $\lambda$. In Fig. \ref{phase1}, we show the phase diagrams for different parameters, we fix $\rho_{0}=1.5$ and $\lambda=3$ but change $\gamma_{0}$ from 0.42 to 0.45.\\ Here, as \cite{Preis:2016fsp}, the chemical potential where the baryonic matter changes to the quark phase, $\mu_{c}$, increases significantly. We find $\mu_{c}$ by calculating free energy for baryonic and quark case, when they are equal a first-order phase transition happens at $\mu_{c}$,
Also we have define table 3, for different parameters, we have found different critical chemical potential.
\begin{table}\label{table4}
\centering
\begin{tabular}{|c| c| c| c| }
\hline
\hline
$\rho_{0}$&$ \gamma_{0}$&$\lambda $&$\mu_{ph}$\\
\hline
2 & 1 & 5 & 21.652 \\
\hline
1.5&.42& 3 & 9.616 \\
\hline
1.5&.45 & 3 & 10.73 \\
\hline
1 & .42 & 3 & 10.19 \\
\hline
\end{tabular}
\caption{We show different phase transition value for different parameters and  non-interaction case.}
\end{table}
\\
\\
\\

\end{document}